\documentclass[12pt]{article}
\usepackage{latexsym}
\usepackage{amssymb}
\usepackage{graphicx}

\def\be{\begin{equation}}
\def\ee{\end{equation}}
\def\bear{\begin{eqnarray}}
\def\eear{\end{eqnarray}}

\newcommand\bra[1]{{\langle {#1}|}}
\newcommand\ket[1]{{|{#1}\rangle}}

\def\a{\alpha}

\def\d{\delta}

\def\th{\theta}

 \def\L{{\Lambda}}
 
 \def\p{\partial}

 \def\IZ{\relax\ifmmode\mathchoice
 {\hbox{\cmss Z\kern-.4em Z}}{\hbox{\cmss Z\kern-.4em Z}}
 {\lower.9pt\hbox{\cmsss Z\kern-.4em Z}}
 {\lower1.2pt\hbox{\cmsss Z\kern-.4em Z}}\else{\cmss Z\kern-.4em Z}\fi}
 \def\IB{\relax{\rm I\kern-.18em B}}
 \def\IC{{\relax\hbox{$\inbar\kern-.3em{\rm C}$}}}
 \def\Ic{{\relax\hbox{$\inbar\kern-.22em{\rm c}$}}}
 \def\ID{\relax{\rm I\kern-.18em D}}
 \def\IE{\relax{\rm I\kern-.18em E}}
 \def\IF{\relax{\rm I\kern-.18em F}}
 \def\IG{\relax\hbox{$\inbar\kern-.3em{\rm G}$}}
 \def\IGa{\relax\hbox{${\rm I}\kern-.18em\Gamma$}}
 \def\IH{\relax{\rm I\kern-.18em H}}
 \def\II{\relax{\rm I\kern-.18em I}}
 \def\IK{\relax{\rm I\kern-.18em K}}
 \def\IP{\relax{\rm I\kern-.18em P}}

 \font\cmss=cmss10 \font\cmsss=cmss10 at 7pt
 \def\IR{\relax{\rm I\kern-.18em R}}

\def\dd{\mbox{d}}

\def\bra{\langle}
\def\ket{\rangle}
\def\a{\alpha}

\def\d{\delta}

\def\G{\Gamma}
\def\e{\epsilon}

\def\f{\phi}
\def\F{\Phi}
\def\vf{\varphi}
\def\k{\kappa}
\def\l{\lambda}
\def\L{\Lambda}
\def\m{\mu}

\def\p{\pi}

\def\th{\theta}

\def\pa{\partial}

\newcommand{\ti}[1]{\tilde{#1}}

\newcommand{\sm}[1]{\mbox{\scriptsize #1}}

\renewcommand{\@}[1]{\sqrt{#1}}
\renewcommand{\le}[1]{\label{#1}\end{eqnarray}}
\newcommand{\bea}{\begin{eqnarray}}
\newcommand{\eea}{\end{eqnarray}}

\newcommand{\eq}[1]{(\ref{#1})}

\def\ffract#1#2{\raise .35 em\hbox{$\scriptstyle#1$}\kern-.25em/
\kern-.2em\lower .22 em \hbox{$\scriptstyle#2$}}

\def\half{{1\over2}\,}
\newdimen\tableauside\tableauside=1.0ex
\newdimen\tableaurule\tableaurule=0.4pt
\newdimen\tableaustep
\def\phantomhrule#1{\hbox{\vbox to0pt{\hrule height\tableaurule width#1\vss}}}
\def\phantomvrule#1{\vbox{\hbox to0pt{\vrule width\tableaurule height#1\hss}}}
\def\sqr{\vbox{%
  \phantomhrule\tableaustep
  \hbox{\phantomvrule\tableaustep\kern\tableaustep\phantomvrule\tableaustep}%
  \hbox{\vbox{\phantomhrule\tableauside}\kern-\tableaurule}}}
\def\squares#1{\hbox{\count0=#1\noindent\loop\sqr
  \advance\count0 by-1 \ifnum\count0>0\repeat}}
\def\tableau#1{\vcenter{\offinterlineskip
  \tableaustep=\tableauside\advance\tableaustep by-\tableaurule
  \kern\normallineskip\hbox
    {\kern\normallineskip\vbox
      {\gettableau#1 0 }%
     \kern\normallineskip\kern\tableaurule}%
  \kern\normallineskip\kern\tableaurule}}
\def\gettableau#1 {\ifnum#1=0\let\next=\null\else
  \squares{#1}\let\next=\gettableau\fi\next}

\makeatletter
\@addtoreset{equation}{section}
\makeatother

\begin{document}

\begin{flushright}
KCL-MTH-07-14\\
\end{flushright}
\vskip1truecm

\centerline{{\Large \bf Holographic Aspects of Electric-Magnetic Dualities}} 
\vskip .5cm
\centerline{{ \bf \Large }}

\vspace{.5cm}

\begin{center}
{\large Sebastian de Haro${}^\star$ and Anastasios C. Petkou${}^\dagger$}\\
\vspace{.5cm}
${}^\star${\it Department of Mathematics\\
King's College, London WC2R 2LS, UK}\\
\tt{sebastian.deharo@gmail.com}\\
\vspace{.2cm}
$^\dagger${\it Department of Physics\\
University of Crete, 71003 Heraklion, Greece}\\
{\tt petkou@physics.uoc.gr}

\end{center}
\vskip 1truecm


\centerline{\textbf{\large \bf Abstract}}
\vspace{.6cm}
We review recent work on holographic aspects of electric-magnetic dualities in theories that involve conformally 
coupled scalars and abelian gauge fields in asymptotically AdS$_4$ spaces. Such models are relevant for the holographic description of M-theory. We also briefly comment on  some new results on
the holographic properties of generalized electric-magnetic duality in gravity.

\section{Introduction and Summary}

The standard AdS$_5$/CFT$_4$ paradigm may be viewed as the holographic image of the decoupling of massive degrees of freedom in  four-dimensional YM theories. In YM theory the decoupling takes place moving from the asymptotically free UV limit to the strongly coupled IR regime. In string theory the decoupling takes place by running down to the low-energy supergavity limit. The field and string theory decoupling limits are  holographically identified, hence supergravity corresponds to strongly coupled field theories. The presence of D-branes, singularities and black holes in string theory is naturally associated to phenomena such as symmetry breaking, confinement and finite-temperature in field theory. 

Recent work on AdS$_4$/CFT$_3$ brought in light a possible new kind of holographic correpondence. Bulk theories that exhibit a generalized form of  electric-magnetic duality correspond to boundary theories whose correlation functions have special transformation properties \cite{Witten, LP}. The latter transformations are induced by certain  ``double-trace" deformations \cite{LP}. This implies the existence of ``duality-related fixed points" in three dimensions, which would be the holographic images of the electric-magnetic dual descriptions of the bulk theories \cite{Tassos1,dHPeng}. Very recently, some intriguing transport properties of three-dimensional theories have also been attributed to bulk electric-magnetic duality \cite{Sachdev,HK,HH}. Such studies, combined with  the observation that three-dimensional field theories have a quite different quantum structure from their four- and two-dimensional counterparts,$\!$\footnote{For a review of three-dimensional theories see \cite{Moshe_Zinn}.} imply that the bulk theory is rather exotic. As electric-magnetic duality appears to be relevant for M-theory compactifications 
it is natural to conjecture that the quantum structure of three-dimensional field theories provides crucial 
information on the non-linear dynamics of M-theory\footnote{A related idea appeared long ago in the context of 
$E_{11}$ \cite{peter}. Here, a non-linear realization of
M-theory based on $E_{11}$ is shown to require the formulation of eleven-dimensional supergravity in terms of
mutually dual fields.}.

In this note we review recent work on various holographic aspects of  generalized electric-magnetic duality and
briefly comment on a few new results concerning gravity. In 
Section 2 we present models with conformally coupled scalars  and in Section 3 the case of a $U(1)$ gauge field 
in asymptotically AdS$_4$. These models can be embedded into M-theory, so our results provide a hint for the role of generalized electric-magnetic duality in the holographic description of the latter. The $U(1)$ gauge field is the first step towards the discussion of YM, gravity and higher-spin gauge fields in AdS$_4$. We summarize and put our findings into perspective in Section 4.


\section{Conformally coupled  scalars}

\subsection{A toy model}

A conformally coupled scalar with quartic self-interaction in fixed Euclidean AdS$_4$ simplifies, after a suitable conformal rescaling of the field and metric, to the model of a massless scalar $\Phi(r,\vec{x})$ with quartic self-interaction $(\lambda/4!)\Phi^4$ on the upper half of $\mathbb{R}_4$ with $r\geq 0$ \cite{dHP}. The boundary is the hyperplane $r=0$. $\F$ behaves near the boundary as
\be
\label{asympt1}
\F(r,\vec{x}) = \alpha(\vec{x}) +r\,\beta(\vec{x}) +\cdots
\ee
with $\alpha$ and $\beta$ arbitrary functions. Requiring regularity of (\ref{asympt1}) for $r\rightarrow \infty$ 
gives a non-local and generally invertible relationship $\beta=\beta[\alpha]$.  It was pointed out in \cite{dHP} 
that  the bulk on-shell action $W_{\sm{ren}}[\alpha]$  of the model evaluated as a functional of $\alpha$ gives 
(minus) the  {\it effective action} of a DBT (Dual Boundary Theory) that contains an operator ${\cal O}_1$ with 
dimension $\Delta=1$. ``Duality" refers here to the dimension of the operator (a duality between CFT's) not to 
the holographic dual. The vacuum structure of the DBT is determined by
\be
\label{NewVac}
\langle{\cal O}_1(\vec{x})\rangle_{\alpha=\a_0} =-\alpha_0(\vec{x}) ~\,,~\,\,\,\,\,\,\frac{\delta W[\alpha]}{\delta\alpha(\vec{x})}\Bigl|_{\alpha_0}=0~.
\ee
For $\lambda<0$ our toy model possesses a non-trivial classical instanton solution
\be
\label{FubLip}
\F(r,\vec{x}) =\sqrt{\frac{48}{-\lambda}}\,\frac{b}{b^2 +r^2 +\vec{x}^2}~\Rightarrow ~\alpha_0(\vec{x})=\sqrt{\frac{48}{-\lambda}}\frac{b}{b^2+\vec{x}^2}~,
\ee
with $b$ an arbitrary parameter with dimensions of length. This implies that the DBT has a non-trivial vacuum structure where the operator ${\cal O}_1$ has non-zero expectation value $-\alpha_0$. A three-dimensional model that may reproduce the bulk results is a massless scalar with self-iteraction $(g/6!)\vf^6$. For $g<0$ this model has the instanton solution
\be
\label{Inst3}
\vf_0(\vec{x}) =\left(\frac{360}{-g}\right)^{1/4}\left(\frac{c}{c^2+\vec{x}^2}\right)^{1/2}\,.
\ee
One can show that the parameters of the bulk and boundary models are related as 
\be
\frac{1}{g} =-\frac{32}{45}\frac{1}{\lambda^2}\,,\,\,\,c=\kappa\, b\,,\,\,\,\,\, \vf_0^2(\kappa\,\vec{x}) =-\langle {\cal O}_1(\vec{x})\rangle \Rightarrow \kappa^2=\frac{16\pi^2}{3\lambda}\,.
\ee
It appears that the bulk and boundary theories in this model are both accessible by standard field theoretic methods. This is connected to the fact that {\it both} $\alpha$ and $\beta$ can be interpreted as expectations values of boundary operators. In a Hamiltonian analysis of the bulk scalar field theory $\alpha$ and $\beta$ play the role of ``coordinate" and ``momenta", hence their mutual interchange resembles the standard canonical transformation $p\rightarrow q$, $q\rightarrow -p$ of Hamiltonian mechanics. This may be termed {\it harmonic oscillator duality} to be contrasted with the usual {\it electric-magnetic duality}. 

To strengthen the above observation we have radially quantized the bulk and boundary theories \cite{dHP}.  The corresponding mode expansions of the  free field configurations are
\bea
\label{radial_quant}
\hat\F(R,\theta,\Omega_2)=\sum_{jlm}\left(\frac{a^{+}_{jlm}}{\sqrt{j+1}}\,R^j\, {\cal Y}^*_{jlm}(\Omega_3)+\frac{a^{-}_{jlm}}{\sqrt{j+1}}\,\frac{1}{R^{j+2}}{\cal Y}_{jlm}(\Omega_3)\right)~,
\\
\hat\vf(R,\Omega_2)=\sum_{\ell m}{1\over\sqrt{2\ell+1}}\left(b_{\ell m}^\dagger\, R^\ell\, Y_{\ell m}^*(\Omega_2)+b_{\ell m}\,{1\over R^{\ell+1}}\,
Y_{\ell m}(\Omega_2)\right)
\eea
with ${\cal Y}_{jlm}$ the hyperspherical harmonics of $S^3$. The boundary is at $\theta=\pi/2$. Our result above implies that the bulk elementary operator is identified with a properly normal ordered boundary composite operator as
\be
\hat\F\left(R,{\pi\over2},\Omega_2\right) \equiv -{\cal O}_1(\vec{x})=:\hat{\vf}^2(x):~.
\ee
The bulk and boundary creation and annihilation operators are related as 
\be\label{abb}
a_{jlm}=\sum_{m_1m_2}c^{\ell\ell m_1m_2}_{lm}\,b_{\ell m_1}b_{\ell m_2}~,~\,\,\,
a_{jlm}^\dagger \sum_{m_1m_2}c^{\ell\ell m_1m_2}_{lm}\,b^\dagger_{\ell m_1}b^\dagger_{\ell m_2}~,~\,\, j=2\ell
\ee
with constant coefficients $c^{\ell\ell m_1m_2}_{lm}$ that were computed in \cite{dHP}. Bulk one-particle states correspond to boundary two-particle states. We have made an effort to extend the above remarkable bulk/boundary quantum correspondence to fluctuations around the non-trivial instanton solutions. 
We obtained a highly non-trivial classical correspondence between the bulk and the square of the boundary 
fluctuations around the instantons\footnote{This result was obtained with T. Koornwinder and 
appears in the appendix of \cite{dHP}.} but we have not yet fully developed the quantum correspondence. 

\subsection{A model embedded in M-theory}

In \cite{dHPP} it was pointed out that conformally coupled scalars also appear in M-theory compactifications. 
Specifically, we considered the model with action 
\be\label{conformal_action}
S=\half\int\dd^4x\sqrt{g}\left(\frac{-R+2\L}{\k^2}+(\pa_\m\f)^2+\frac{1}{6}R\f^2+\l\f^4\right)
\ee
where $\k^2=8\p G_4$, $\l$ is a dimensionless coupling and the cosmological constant is $\L=-3/l^2$.
For the special value of the quartic coupling $\l=\k^2/6l^2>0$ the action (\ref{conformal_action}) is obtained by a consistent truncation of 11-dimensional supergravity. 

The asymptotic behavior of the scalar $\f$ is as in (\ref{asympt1}). Motivated by the existence of the instanton solution (\ref{FubLip}) we can impose the ``self-dual" boundary conditions $\beta=-la\alpha^2$ by adding the boundary term
\be\label{b_term}
S_{\sm{bdy}}=-\frac{l^3 a}{3}\int\dd^3x\,\alpha^3(\vec{x})~.
\ee
We consider solutions with vanishing energy momentum tensor in which case the background is still AdS, since all extrema of the action (\ref{conformal_action}) have constant Ricci scalar $R=-12/l^2$ \cite{IP}. It is quite remarkable that in this case we are able to calculate {\it exactly} the effective potential (constant $\alpha$) of the DBT as
\be\label{eff_potential}
V_{\l,a}(\a)=\frac{1}{3\l}\left[\left(\frac{R}{6}+\l\a^2\right)^{3/2}-a\l\a^3
-\left(\frac{R}{6}\right)^{3/2}\right],
\ee
as well as its effective action in the double-scaling limit $\lambda -a^2=\mu \rightarrow 0$ 
\be\label{lim_eff_action}
\G_{\sm{eff}}[\alpha]=\frac{1}{3a}\int d^3x\sqrt{g_{(0)}} \left(\frac{1}{2}\pa_i\vf \pa^i\vf +\frac{1}{16} R[g_{(0)}]\vf^2 +\frac{1}{8}\m\vf^6\right)+{\cal O}\left(a^{-2}\right),
\ee
where  $\vf^2=\a$ and $g_{(0)ij}$ is the boundary metric. Remarkably, (\ref{lim_eff_action}) coincides with the conformal three-dimensional models used in \cite{dHP}. 

For $\lambda>0$ the action (\ref{conformal_action}) has the instanton solution 
\be\label{instanton_uhp}
\f=\frac{2}{l\sqrt{|\l|}}\left(\frac{Br}{-{\rm sgn}(\l)B^2
+(r+A)^2+(\vec{x}-\vec{x}_0)^2}\right),
\ee
where $A,B,x_0^i$, $i=1,2,3$, are arbitrary constants. This is non-singular provided $A>B\geq 0$ and satisfies the ``self-dual" boundary condition. The existence of this solution is rather surprising and implies the instability (a la Coleman-de Luccia) of pure AdS$_4$ towards the spontaneous dressing by  a scalar field. In also implies the possible instability of a stack of M2-branes. The decay rate of the vacuum is 
\be\label{decay_rate}
{\cal P}\propto \exp(-\Gamma_{\sm{eff}}\bigl|_{\sm{inst}})\,,\,\,\,\left.\G_{\sm{eff}}\right|_{\sm{inst}}=\frac{4\pi^2l^2}{\k^2}\left(\frac{1}{\sqrt{1-\k^2/6l^2 a^2}}-1\right),
\ee
with $a=\sqrt{|\lambda|}A/B$. Note that the deformation parameter $a$ drives the theory from the regime of marginal
stability at $a=\kappa/\sqrt{6}l$ to total instability at $a\to\infty$.

\section{Abelian gauge fields}

This is the first instance where we encounter standard electric-magnetic duality in the bulk of asymptotically AdS$_4$. We have 
the freedom to identify the leading or subleading terms in the gauge field as the source or the operator of the boundary theory. This is
the ``generalized electric-magnetic duality". But for gauge fields we can also choose between ``electric" and ``magnetic" boundary
conditions. This choice corresponds to adding a marginal, abelian $AB$-type Chern-Simons term on the 
boundary  \cite{Witten,HUY}.

We will first discuss regular bulk instanton solutions. These are special because they provide exact solutions of
the coupled gravity-matter field equations whilst allowing for non-trivial dynamics of the gauge field at the 
boundary \cite{dHPeng}. 
Instantons are regular self-dual solutions $F=*F$ of the Euclidean field equations. The regularity condition 
in the interior relates the boundary value of the electric field to the 
boundary value of the transverse part of the gauge field:
\be
E_i(p)=-{1\over g^2}\,|p|\,A_i(p)~.
\ee
The on-shell effective action now gives the generating functional of the boundary theory:
\be
W[A]=-\half\int\dd^3x\,A_i(x)E_i(x)
={1\over2g^2}\int\dd^3p\,|p|\,A_i(p)A_i(-p)~.
\ee
The partition function is purely a functional of $A_i$, which is interpreted as a source in the CFT. The above
is consistent with a Dirichlet choice of boundary conditions where $\d A_i=0$ at the boundary. 
However, gauge fields in AdS$_4$ admit a more general choice of boundary conditions, for instance those
corresponding to bulk instantons. To get a variational problem that generates instanton boundary conditions we 
need to modify the action by a boundary Chern-Simons term:
$S_{\sm{CS}}={\th\over8\pi^2}\int\dd^3x\,\e^{ijk}A_i\pa_jA_k$. The boundary conditions are now modified to
\be\label{abelianeq}
\Box^{1/2}\,A_i+{\th g^2\over4\pi^2}\,\e_{ijk}\pa_jA_k=0~.
\ee
For $\th=\pm4\pi^2/g^2$, this gives regular bulk instanton solutions. As noticed in \cite{dHPeng}, this is the
equation of motion for a topologically massive spin-1 particle in three dimensions \cite{PTvN,DJ}, with mass 
$m\sim \Box^{1/2}$. {\it Any} regular, self-dual solution satisfies \eq{abelianeq}
with the corresponding value of the theta angle \cite{dHPeng}.

To discuss $S$-duality, we go back to generic solutions of the Dirichlet boundary problem. Holographically, $A_i$ 
is a source that couples to a conserved current of dimension 2. The two-point function of this current
is computed to give the familiar result, plus a parity-breaking term that comes from the theta angle:
\be
\bra J_i(p)J_j(-p)\ket={1\over g^2}\,|p|\,\Pi_{ij}+{i\th\over4\pi^2}\,\e_{ijk}p_k~.
\ee
$\Pi_{ij}$ is the projector onto transverse vectors. For simplicity we set $g=1$ and $\th=0$, the general case being discussed
in \cite{LP,dHPeng}. 

The conserved current $J_i$ is equal to the boundary value of the
electric field. The bulk equations have electric-magnetic invariance which interchanges the boundary
values of the electric and magnetic fields: $B_i\leftrightarrow\pm E_i$. However, the boundary terms are not
invariant; the Dirichlet and Neumann boundary problems get interchanged \cite{Witten}. 

Having seen the Dirichlet problem, let us now discuss the dual Neumann quantization scheme. Now the electric field is held fixed at
the boundary whereas the magnetic field is fluctuating. Hence, the path integral formulation
includes an integral over boundary configurations of the gauge field $A_i(x)$ up to gauge transformations.
The holographic interpretation of this is that the electric field corresponds to a dual source, and $A_i$ to
an operator. There is a subtlety though. $A_i$ itself cannot be the dual operator because it has dimension 1 and 
is below the unitarity bound. However, we can construct from it a dual current $\ti J=*\dd A$ with dimension 2
\cite{LP,dHPeng}. This current corresponds to the bulk magnetic field. It is conserved and lies on the marginal line of the 
unitarity bound, hence its correlators correspond to a unitary theory, the DBT. Further,
since the electric field is conserved, it can be written in terms of a 1-form $J=*\dd \ti A$. This 1-form is the
one that is now fixed, and is identified with the dual background field \cite{Witten}. 

In summary, bulk electric-magnetic duality
interchanges $A\leftrightarrow\pm\ti A$ on the boundary. The Dirichlet problem corresponds to fixing the
gauge field on the boundary, with the electric field corresponding to a dimension-2 current. In the Neumann 
problem, the magnetic field corresponds to the dual current $\ti J=B$ whereas the dual gauge field is the
source. The dual generating functional is $\ti W[\ti A]=\half\int\dd^3x\,\ti A_iB_i$.
Both theories are related by a Legendre transform with an $AB$-type Chern-Simons term $-\int A\wedge\dd\ti A$
\cite{Witten,dHPeng}. 

The embedding of this model in eleven-dimensional supergravity was discussed in \cite{dHPeng,Sachdev}.

\section{Gravity and M-theory holography}

Gravity is known to exhibit similar electric-magnetic properties to the ones
discussed here \cite{LP2}\footnote{See also \cite{peter} and references in \cite{dHPeng}.}. Whereas we will discuss the implications
for duality in the CFT elsewhere
\cite{toappear}, we will give some new results here. Despite some profound differences with the lower spin cases,
there is evidence that duality in the CFT works in a similar way for gravity. Namely, not only do we have the possibility 
to choose between leading and subleading terms in the graviton expansion, but we also can choose 
between ``electric" and ``magnetic" boundary conditions. The latter corresponds to adding a gravitational 
Chern-Simons term on the boundary. 

We observe a pattern in the M-theory compactifications. Firstly, the three-dimensional boundary theories do not exhibit the decoupling of massive modes that we observe in four-dimensional YM theories. Namely, the theories related by ``double-trace" deformations correspond to fixed-points which appear to have very similar operator content, in contrast with YM theories where the weak and strong coupling limits have very different content. Secondly, the Chern-Simons terms in the boundary should correspond to parity anomalies of three-dimensional theories. One may conjecture that the structure of three-dimensional theories is a hologram of the structure of the full M-theory. Namely, we expect that full blown M-theory is duality invariant, however only after its full spectrum is taken into account. This duality, which is a generalization of electric-magnetic duality, is broken at the level of 11-dimensional supergavity, but we see a sign of it in the boundary conditions. 

To end we will make some remarks on the generalization of the results in the previous section to the gravitational
field. We will consider instanton configurations where the Weyl tensor is self-dual. It can 
be shown \cite{toappear} that the stress-energy tensor of in that case is given by
the Cotton tensor,
\be
\bra T_{ij}\ket={\ell^2\over8\pi G_N}\,C_{ij}~,
\ee
computed with respect to a representative of the conformal class of boundary metrics $g_{(0)ij}=g_{ij}(r=0,x)$ 
obtained from the bulk (for notation, see \cite{SdHSSKS}). Recall
that the Cotton tensor is symmetric, traceless, and conserved. According
to \cite{SdHSSKS} the above is a boundary condition for the third derivative of the rescaled metric $g_{ij}(r,x)$, $g_{(3)}$. 
This result is generic and does not depend on a particular choice of asymptotics $g_{(0)}$ for the bulk metric.
Expanding around an AdS$_4$ background, $g_{ij}(r,x)=\d_{ij}+h_{ij}(r,x)$, we look for regular instantons. Using the
results in \cite{GlebSergei}, one can show that regularity imposes
\be
h_{(3)ij}={1\over3}\,\Box^{3/2}h_{(0)ij}~,
\ee
and we have projected onto the transverse, traceless part of the graviton. Combining both conditions above, we get
that regular bulk instantons satisfy
\be\label{SDgrav}
\Box h_{(0)ij}=\a\, \e_{ikl}\Box^{1/2}\pa_k h_{(0)jl}+(i\leftrightarrow j)~.
\ee
This is a non-trivial differential equation for the boundary metric that generalizes the {\it self-dual} boundary conditions of the scalar and the $U(1)$ case \eq{abelianeq}.
We stress that {\it any} self-dual, regular bulk solution expanded about AdS$_4$ satisfies the above condition
\eq{SDgrav}.

\section*{Acknowledgements}

SdH thanks the organizers of the EPS conference HEP2007 in Manchester, where this talk was delivered. The
research of SdH is supported by PPARC and EU grants PP/C507145/1, MRTN-CT-2004-512194.
The work of A.~C.~P. was partially supported by the program
``PYTHAGORAS II" with KA 2101, of the Greek Ministry of Higher Education.

\end{document}